# EMERGING WIRELESS TECHNOLOGIES IN THE INTERNET OF THINGS: A COMPARATIVE STUDY

Mahmoud Elkhodr, Seyed Shahrestani and Hon Cheung

School of Computing, Engineering and Mathematics, Western Sydney University, Sydney, Australia

## ABSTRACT

*The Internet of Things (IoT) incorporates multiple long-range, short-range, and personal area wireless networks and technologies into the designs of IoT applications. This enables numerous business opportunities in fields as diverse as e-health, smart cities, smart homes, among many others. This research analyses some of the major evolving and enabling wireless technologies in the IoT. Particularly, it focuses on ZigBee, 6LoWPAN, Bluetooth Low Energy, LoRa, and the different versions of Wi-Fi including the recent IEEE 802.11ah protocol. The studies evaluate the capabilities and behaviours of these technologies regarding various metrics including the data range and rate, network size, RF Channels and Bandwidth, and power consumption. It is concluded that there is a need to develop a multifaceted technology approach to enable interoperable and secure communications in the IoT.*

## 1. INTRODUCTION

The Internet constitutes the largest heterogeneous network and infrastructure in existence. It is estimated that over 3 billion people had access to the Internet in 2014. Also, there are as many mobile subscriptions (6.8 billion) as there are people on earth [1]. Global mobile data traffic was estimated at 2.5Exabyte per month in 2014 [2]. This figure is estimated to rise to 24.3 Exabyte per month at a compound annual growth rate of 57 percent in 2019 [3]. This can be attributed to a number of technological factors including the proliferation of touch screen devices (smartphones, tablets, and the like), and, significantly, the evolvement and technological advancement of wireless and mobile technologies. On the other hand, the Internet of Things (IoT) is a fast-growing heterogeneous network of connected sensors and actuators attached to a wide variety of everyday objects. Mobile and wireless technologies in their assortment of low, ultra-power, short and long range technologies continue to drive the progress of communications and connectivity in the IoT. The future will foresee smart and low-power networked devices connecting to each other and to the Internet using, mostly, reliable low-power wireless transmissions. Figure 1 shows the rapid growth of IoT by 2020.

This paper investigates and compares some of the evolving and enabling wireless technologies for the IoT. It analyses the capabilities of IEEE 802.15.4 technologies, Bluetooth Low Energy, and Wi-Fi. Additionally, it explores the opportunities promised by the recent development in IEEE 802.11ah and LoRa technologies. LoRaWAN and IEEE 802.11ah are the latest technologies in long-range and low-power WAN. They are targeted for low-power and low-cost devices. LoRaWAN targets key requirements of the IoT such as secure bi-directional communications, mobility, and localization services. This standard will provide seamless interoperability among smart things without the need of complex local installations, and gives back the freedom to the users, developers, and businesses aiding the flourishment of the IoT. For instance, LoRa plays a significant role in the future of wireless and machine to machine (M2M) communications. On the other hand, 802.11ah is IEEE latest update to their legacy 802.11 technologies (popularly known as Wi-Fi). IEEE 802.11ah aims to cater for low-cost and low-power market. It is a competitor to LoRa, ZigBee and other technologies in their class.

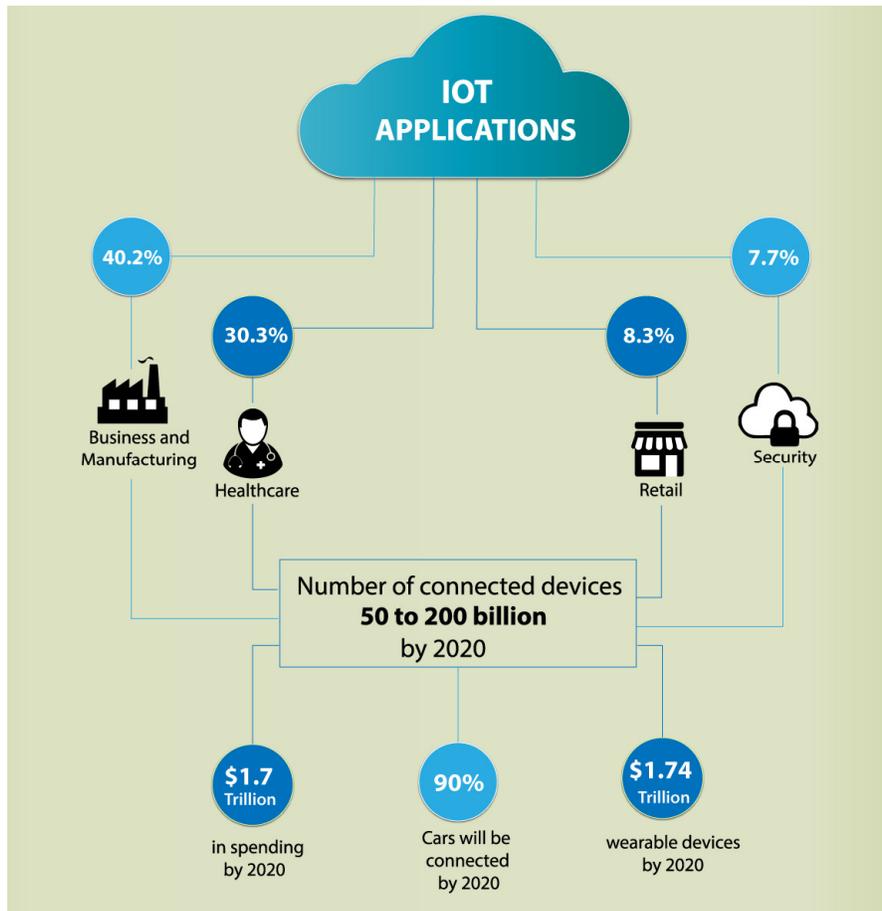

Figure 1- IoT Growth by 2020

## 2. WIRELESS LOW-POWER TECHNOLOGIES FOR THE IoT

The IoT covers a broad range of applications and devices. The 802.11 protocol with its 802.11a/b/g/n/ac variants is among the first obvious technology candidates for the IoT. Examples of Wi-Fi applications in the IoT are presented in [4]. Today, almost every house, workplace, cafe, and university has a Wi-Fi network. Wi-Fi has become the de-facto term when referring to connecting to the Internet via a wireless access point. The widespread adoption of Wi-Fi makes it a first technology choice for many IoT applications. However, in some IoT applications, the choice of technology is limited to the devices hardware capabilities, low-power consumption requirements, and the overall cost. Many IoT devices require the use of a low-cost and low-power wireless technology when connecting to the Internet [5]. Traditionally, energy consumption has always been a limiting factor in many wireless sensor network applications. This limiting factor will continue as a major challenge facing the development of many applications in the IoT. In fact, for the growth of the IoT, low-power consumption is an essential requirement that needs to be met.

In addition to low-power consumption, other associated requirements need to be considered as well. For instance, the cost of technology, security, simplicity (easy to use and manage), wireless data rates and ranges, among others, such as those reported in [6], are essential requirements that require attention. Many evolving wireless technologies such as ZigBee and Bluetooth are competing to provide the IoT with a low-power wireless connectivity solution. Other wireless technologies such as the IEEE 802.11ah, LoRa, and 6Lowpan protocols are emerging as well [7]. They offer similar low-power wireless connectivity solutions for the IoT. Consequently, there could be many choices of low-power wireless protocols for many IoT applications. Consider, for example, a car-parking system application based on the IoT such as the one presented in [8]. The IoT-based car parking system combines many components together. It combines a variety of devices, multiple networking protocols, several sources of data, and various wireless and generations of technologies. Many of the devices involved in the communications are lightweight devices such as sensors that operate on batteries. They would require a low-power wireless technology to function effectively.

Essentially, low-power wireless technologies contribute to improving not only the way an IoT device connects to the Internet but the efficiency of the IoT application operation as well. A network consisting of low-cost and lightweight IoT devices can be used to monitor relevant operation and contextual parameters. These devices are also capable of making appropriate decisions (based on the occurrence of specific events) while simultaneously communicating with some other IoT devices. In general, a heterogeneous setup allows an IoT system to perform many automated tasks by combining the various data gathered from these IoT devices. In the smart home IoT application example, IoT devices such as wireless sensors can report the ambiance temperatures in various locations in a house to an IoT central device, referred to as the controller, which in turns can make a decision on varying the output of the air-conditioning system. Adding more IoT devices to the IoT system will increase the intelligence of the system as well. For instance, if some other sensors are providing information on whether the house is occupied or not (whether the people occupying the house are out or no), then the controller will be able to make a better decision on when the heating system should be turned on or off. In this smart home example, the IoT devices are in the form of simple sensor devices which have a small bandwidth and low-power requirement. Hence, the need for low-power wireless technologies in this and many other similar applications in the IoT is essential. Other WSN applications in the IoT feature the use of sensor devices that monitor critical infrastructure or carry sensitive information. They are often deployed in remote areas in an ad-hoc fashion, which raises many challenges. Table 1 briefly lists some of the major issues challenging the incorporation of WSNs in the IoT.

## 2.1 Analysis of IEEE 802.11 WLANs for IoT Communications

Wireless Local Area Networks (WLANs) is the dominant technology for indoor broadband wireless access. WLAN products have become commodity items used in professional and consumer products alike. Recently, the propagation of WLANs as extensions of wired networks has been increasing dramatically, and thereby, giving devices equipped with wireless interfaces a higher degree of mobility. The two most common WLAN standards are the IEEE 802.11 standard (commonly branded as Wi-Fi) and the European HIPER (HighPerformance Radio) LAN [9]. The IEEE 802.11 defines two types of configurations, the Infrastructure Basic Service Set (iBSS) and Independent BSS (IBSS). In iBSS, an access point (AP) is the central entity of each coverage area with coordination functionality. Additionally, the AP acts as a bidirectional bridge between the wireless network and the wired infrastructure (i.e., typically Ethernet). Stations (STA) are mostly mobile devices equipped with IEEE 802.11 wireless network interfaces. Communication between the AP and the associated stations occurs over the shared wireless medium that carries the data. A station must associate with an AP for it to transmit and receive data to and from the wired infrastructure, and to communicate with other stations on the same WLAN. A Basic Service Set (BSS) is the term used to refer an AP and its associated stations. In large WLANs, multiple BSSs can be joined using a distribution system (DS), thus providing sufficient coverage for a greater number of stations. This setup of having two or more BSSs is referred to as an Extended Service Set (ESS). The DS is the wired backbone connecting APs and allowing the associated stations to access services available on the wired infrastructure.

Therefore, Wi-Fi devices can form a star topology with its AP acting as an Internet gateway. The output power of Wi-Fi is higher than other local area network wireless technologies. Full coverage of Internet connectivity is necessary for Wi-Fi networks, so dead spots which may occur are overcome by the use of more than one antenna in the AP. Wi-Fi operates in the 2.4 and 5 GHz bands. Its operations in the 5 GHz band allow the use of more channels and provide higher data rates. However, the range of 5 GHz radio indoors (e.g., inside buildings) is shorter than 2.4 GHz. The IEEE 802.11b and IEEE 802.11g operate in the 2.4 GHz ISM band. The IEEE 802.11n improves the previous versions of the standard by introducing the multiple input and multiple output methods (MIMO) [10]. It supports a data rate ranging from 54 Mbit/s to 600 Mbit/s [11]. The IEEE 802.11ac is an improved version of the IEEE 802.11n, and it provides high throughput wireless local area networks (WLANs) in the 5 GHz band with more spatial streams and higher modulation with MIMO yielding data rates up to 433.33 Mbps [12].

The IEEE 802.11ac provides a single link throughput of at least 500 Mbps and up to 1 gigabit per second. The IEEE 802.11ac has a wider RF bandwidth of up to 160 MHz and a higher density modulation up to 256 QAM [13]. At the other end of the spectrum, the IEEE 802.11ah standard operates in the unlicensed 900MHz frequency band. A wireless signal operating in the 900MHz band can penetrate walls, but it would deliver a limited bandwidth ranging from 100Kbps to 40Mbps [14]. One common IoT application of this technology would be sensors and actuators in homes or commercial buildings. Thus, IEEE 802.11ah could be positioned as a competitor to Bluetooth and ZigBee protocols in the IoT space.

Table 1- Some Challenges of IoT WSNs

| Challenges | Description |
|---|---|
| Energy | IoT sensor devices are typically powered by batteries.There is a need to eliminate redundant data or aggregate sensor readings |
| Self-Management | Ad-hoc deployment of IoT sensors (many sensor networks are deployed without design) |
| Unattended Operation | self-organization, self-optimization, self-protection, and self-healing |
| Multi-hop Communication | signal attenuation, increased latency, radio range etc. |
| Design Constraints | Lack of I/O components such as GPS receivers, Centralized vs decentralised management, |
| Security | Conventional security techniques often not feasible due to their computational, communication, and storage requirements |

## 2.2 IEEE 802.11ah, ZigBee IP and Bluetooth Low Energy

Wi-Fi, with its 802.11 a/b/g/n/ac variants, may not be deemed suitable to use in some IoT applications where low-power consumption is a vital requirement. Wi-Fi was originally designed to offer high throughput to a limited number of devices located indoor at a short distance from each other. Therefore, to meet the IoT low-power requirements, the IEEE 802 LAN/MAN Standards Committee (LMSC) formed the IEEE 802.11ah Task Group (TGah) [15]. The task group objective is to extend the applicability area of 802.11 networks. It aims to design an energy efficient protocol allowing thousands of indoor and outdoor devices to work in the same area [16]. The IEEE 802.11ah seeks to support a range of throughput options ranging from 150 Kbps up to 40 Mbps over an 8 MHz band [14]. With regard to the wireless range, the proposed IEEE 802.11ah protocol supports a wider coverage range when compared to that of the IEEE 802.11n/ac protocol. The IEEE 802.11ah supports applications with coverage of up to 1 km in outdoor areas and up to 8191 devices associated with one access point [17]. The IEEE 802.11ah operates in the unlicensed sub-1GHz bands, excluding the TV white-space bands. Sub 1 band provides an extended wireless range when compared to the other bands used by conventional 802.11 Wi-Fi standards which operate in the 2.4 GHz and 5 GHz bands [18]. The IEEE 802.11ah relies on the one-hop network topology and employs power saving mechanisms [14]. Given that the IEEE 802.11ah protocol falls under the overall WiFi umbrella, it is expected that it will be compatible with the existing Wi-Fi infrastructure [19]. The IEEE 802.11ah allows access to more than 8 thousand devices in the range of 1 km within an area with high concentration of small devices such as sensors, and mini controllers. Therefore, the IEEE 802.11ah technology can satisfy the IoT requirements while maintaining an acceptable user experience in parallel with the IEEE 802.11 technologies. One of the interesting functional requirements of the IEEE 802.11ah is to enable coexistence with the IEEE 802.15.4 standard [20].

The IEEE 802.11ah standard includes new PHY and MAC layers grouping devices into traffic induction maps to accommodate small units and machine to machine (M2M) communications [21]. The physical layer allows devices along with the AP to operate over various sub-1GHz ISM bands depending on the regulation of the country [21]. The 900 MHz band is currently used in Europe for GSM 2G cellular facilities. The 900 MHz is used in many devices, and it is suitable forM2M communications specifically in constrained devices such as wireless sensors. In some countries, the frequency bands vary from 902-928 MHz in the USA, 863-868.6 in Europe, 950.8-957.6 MHz in Japan. Other countries are expected to follow in releasing the spectrum once the IEEE 802.11ah standard is finalised. On the other hand, the IEEE 802.11af also called Super Wi-Fi or White-Fi, operates in the unused TV spectrum [22]. 802.11af coverage can extend up to several kilometres as it operates in the frequency bands between 54MHz and 790MHz. It offers a reasonable throughput, estimated at 24Mb/s. It has similar applications as 802.11ah, providing bandwidth for sensors and other devices of the IoT [23].

On the other hand, Bluetooth Low Energy (BLE) 4 is an enhancement to the classic Bluetooth protocol [24]. The low-power consumption feature of the BLE protocol enables connectivity, monitoring, and sharing of information for many devices, such as home appliances and wearable devices, with a minimal consumption of energy. Significantly, the BLE protocol creates opportunities for a number of IoT applications. It is a strong candidate to be used as a communication protocol in several IoT devices which are limited by their low-power and low-cost characteristics. Examples of these IoT applications and devices range from health monitor devices in e-health, devices in retails applications and home automation systems [25], and smart appliances in smart grid applications [26]. Additionally, the widespread adoption of smartphones and the advancements made by BLE with regard to energy consumption enabled the introduction of many wearables and fitness devices that integrate with smartphones. BLE has a real potential for becoming an essential technology for the last 100 meters in low-power and low-cost small devices of the IoT [25]. Using a smartphone or another similar device as a temporary or mobile gateway is increasingly getting popular in numerous IoT applications. Thus, BLE plays a significant role in providing the communication medium needed between this gateway and the IoT devices. The BLE protocol is designed to have an over the air data rate of 1 Mbps and throughput of around 0.27 Mbps [27]. Similarly, an improvement to the standard ZigBee is ZigBee IP or Smart. ZigBee IP incorporates technologies, such as 6LoWPAN [28], that optimise routing and meshing in wireless sensor networks. It supports the requirements of ZigBee Smart Energy as well [29]. This combination of technologies offers a solution that enables the extension of IEEE 802.15.4 based networks to IP-based networks. ZigBee has a data rate that ranges from 20 to 250 kbps [30]. It operates in the unlicensed 2.4 GHz band which overlaps with other wireless technologies (e.g., Wi-Fi and Bluetooth) sharing the same band. ZigBee-IP provides a scalable architecture that supports an end-to-end networking based on IPv6. Therefore, many applications in the IoT benefit from this architecture [31].

## 2.3 LoRaWAN

As seen in the previous section, technologies such as ZigBee, IEEE802.11ah, and BLE provide low-power solutions for many IoT applications. However, these technologies suffer from major shortcoming in regards to coverage. Their shortrange communications limitations make their deployment in many smart city and remote IoT applications unfeasible. In the IoT, several IoT devices require to transmit data over longer distances while running on batteries. Sensors and actuators in industrial IoT applications requires low bandwidth communication over longer distances in condensed indoors and outdoors areas as well. Low power wide area (LPWA) technology has been introduced specifically to deal with short-range limitations of these low-power technologies. While ZigBee and IEEE802.11ah can extend their coverage using meshing technologies, low-power, wide-area network (LPWAN) eliminates many overhead associated with the use of meshing such as forwarding and routing overheads.

Consequently, LPWANs are projected to provide connectivity to numerous devices in several IoT applications. LoRa is an LPWAN wireless technology that provides a wide area network capability. It is often referred to as LoRaWAN. The technology has been developed to support low-power communications over long distances. A single LoRa gateway can cover an entire city similar to that of a cellular network cell [32]. Depending on the obstructions and physical characteristics of an environment, LoRa can cover hundreds of square kilometres. In contrast to cellular technologies that support high data throughput, LoRa is designed for IoT devices and M2M applications that require the exchange of only small amounts of data over longer distances. However, LoRa provisions multiyear battery lifetime as opposed to few hours or days with the use of Cellular technologies. A typical LoRa network consists of the followings entities:

- **LoRa end devices**: These are the endpoint IoT devices that do the sensing or actuation
- **LoRa gateways**: Similar to an IEEE 80211ah gateway or a ZigBee/6lowpna coordinator, a LoRa gateway receives communications from LoRa end devices and offers Internet backhaul functionality. LoRa gateways are projected to be housed with cellular base stations.
- **LoRa server**: It is a network server that manages the LoRa network including packets filtration, data rate adaptation among many other network managements and control capabilities.
- **LoRa remote computer/cloud system**: Provides high level application services such as collecting and processing the data gathered by end devices, performing data analytics, and running IoT applications.

LoRaWAN typically implements a star network topology whereas gateways relay messages between end-devices and the LoRa server. All end-point communication is generally bi-directional with data rates ranging from 0.3 kbps to 50 kbps. LoRa adopts an adaptive data rate (ADR) scheme. This scheme allows the LoRa server to manage the data rate and RF output of individual nodes. This helps in optimizing the battery consumptions of LoRa end devices each according to its application requirements. The nodes in a LoRaWAN network transmit data using the Aloha method. That is, communications are asynchronous in which end-devices send data based on an event-driven or scheduled approach. LoRaWAN defines three classes of end devices [32]:

1) **Class A devices:** Class A is the lowest power end-device as it is most energy efficient. Class A devices spend most of their time in sleep mode. They only wake at a scheduled time or when they are ready to transmit data (event-driven). Thus, communications from the server are only possible during the scheduled uplink of an enddevice.

2) **Class B devices:** Class B devices receive a timesynchronized beacon from the LoRa gateway which allow them to open extra receive windows at scheduled times. This allows the server to determine the time an end-device can receive data.

3) **Class C devices:** Class C end-devices are able to receive data at any time, except when transmitting data, as they are always listening. Obviously, this is the lowest energy efficient class of devices.

# 3. ON THE ADOPTION OF WIRELESS TECHNOLOGIES IN THE IoT

There is a growing momentum to embrace and design technologies that adhere explicitly to the IoT requirements. This includes the modification of existing technologies e.g., from Bluetooth classic to Bluetooth smart, and from ZigBee classic to ZigBee-IP or the design of new technologies such as the IEEE 802.11ah. These technologies aim at addressing key IoT wireless and devices requirements such as low-power consumption, lower computation capabilities, reduced implementation and operational costs, and a wider coverage range. The previous section provided a brief review of the IEEE 802.15.4 technologies, Bluetooth Low Energy, and the IEEE 802.11ah technology. The IEEE 802.15.4 family of technologies, such as the 6Lowpan and ZigBee technologies, are currently used in various wireless sensor network applications. These applications are characterised by requirements similar to those encountered in the IoT. For instance, BLE is widely adopted in wearables and consumer products. On the other hand, the IEEE 802.11ah is a new protocol under development. It is designed to operate in the sub-one-gigahertz (900MHz) band. It has an extended range when compared to traditional Wi-Fi, and it is regarded as a competitor for both ZigBee, 6Lowpan, and the other already-established protocols in this sub-one band.

However, all the technologies above have their weaknesses and obviously their strengths. For example, the gain in range with the use of the IEEE 802.11ah is lost in bandwidth. Whereas, with the use of ZigBee the gain in bandwidth is lost in range.

The areas of the IoT involve diverse sets of devices that use various communication technologies to share and exchange information. Within the IoT, some applications can be in the form of simple peer-to-peer applications. Other IoT applications can also be based on personal area network setups, involving the use of few devices and users. Other complex applications may involve the utilisation of a variety of heterogeneous devices which communicate using a wide array of technologies, in different setups and topologies. Therefore, a technology that can be deemed suitable for a particular IoT application might not necessarily be suited to many others. In fact, the ability to connect and coexist various devices operating using several communication technologies is the vision behind the IoT. Having an ecosystem of coexisted technologies and devices is what enables the IoT vision of extending communications to anything and anywhere.

## 3.1 WLANs: Capacity vs. IoT Requirements

The IEEE 802.11ac and LTE Advanced have the highest data rate among the wireless technologies in use today. The IEEE 802.11ac specification provides a theoretical maximum data transfer speed of more than 3Gbps. It can provide a transfer speed up to 1.3Gbps as well, and supports up to 8 streams [33]. On the other hand, LTE Advanced has a 1Gbps fixed speed and a rate of 100 Mbps to mobile users [34]. Figure 1 compares between various wireless technologies regarding distance coverage in meters, rates, ranges, and power consumptions. In the low-power wireless technology space, Bluetooth Low Energy has the highest data rate of 2.1 Mbps.ZigBee and 6Lowpan technologies, supported by the IEEE802.15.4 standard, have a data rate of 250 Kbps in the 2.4GHz frequency band.

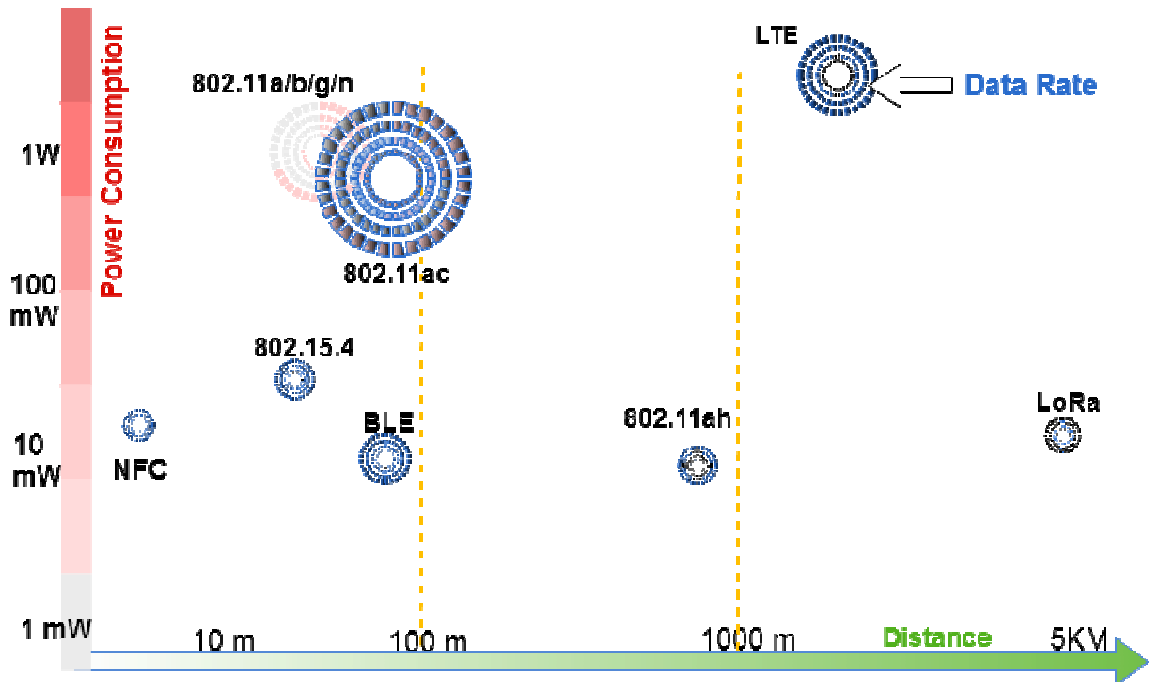

Figure 2- A Comparative Study of Power Consumption, Distance Coverage in Meters, and Data Rate.

However, ZigBee data rate falls to 20 Kbps in the 868 MHz band and to 40 Kbps in the 915 MHz band in some countries [35]. In contrast, the IEEE 802.11ah has the lowest data rate targeted at 150Kbps with an average of 100 Kbps. As of the theoretical wireless range, as illustrated in Figure 2, cellular technologies, e.g., LTE, cover a larger area when compared with other Wi-Fi technologies with IEEE 802.11 variants coming second at an approximate maximum range of a 100 m.

As of the range of low-power wireless technologies, the IEEE 802.11ah rules the chart against 802.15.4 and BLE technologies in the short-range space. The 802.11ah coverage range also outperforms that of the other variants of the 802.11 protocol, with a range coverage of approximately 1 km, as shown in Figure 2 and 3. It should be noted that the 802.15.4 supports mesh networking. In mesh networking, a message is routed through several nodes on a network until it reaches its destination. Therefore, a ZigBee networks range can be easily extended with the use of repeaters in a mesh formation. Data in a ZigBee network hops around a mesh of nodes until a route to the host (usually the Internet) is found. Therefore, repeaters and/or a high density of nodes can be used to extend the coverage of a ZigBee network. Interestingly, the IEEE 802.11ah is under development with meshing in mind as well. On the other hand, LoRa is promising to have a coverage similar to that of cellular networks. This constitutes a huge jump in range coverage. However, this would not be possible without involving a LoRa service provider, which may raise the associated cost derived from the use of this technology.

Therefore, the choice of technology regarding data rate and range come back to the requirements of the IoT applications in hand. Accordingly, if an IoT application requires the use of a larger number of nodes and meshing is an option, ZigBee appears to be a suitable candidate given its data rate advantage over its 802.11ah counterpart. On the other hand, IoT applications that require the deployment of fewer nodes with minimal traffic, 802.11ah is a strong contender to ZigBee. This is because 802.11ah has a larger coverage area without relying on any meshing technique. Also, it is intended to be backward compatible with the variants of 802.11 Wi-Fi technologies. However, as we will see in the next subsections that the data rate and range parameters do not provide enough and sufficient measures when comparing IoT wireless technologies as other criteria need to be considered as well.

## 3.2 Network Size Capabilities for IoT Networks

The BLE protocol supports a maximum of eight nodes per network which include one master device and seven devices as slaves. ZigBee can have up to 65,000 nodes per network in a star topology [30]. These technologies can be extended to more sophisticated networks as well. For instance, ZigBee can be extended to a cluster tree or mesh network; while BLE can be extended to a scatternet network. An interconnected piconet consisting of more than eight Bluetooth devices is referred to as a scatternet. It is the process of connecting two piconets together. A scatternet can be created when a device belonging to one piconet is elected to be a member of the second piconet as well [36]. On the other hand, the baseline IEEE 802.11 standard does not limit the number of devices in the network. However, the limitation can be attributed to the length of some of the fields defined in the management frames of the standard [37]. The Association Identifier (AID) which is a unique value assigned to a station by the AP during an association handshake, is 14 bits long. However, the values other than 1-2007, which are 0 and 2008-16383, are reserved. In particular, AID of value 0 is reserved for group addressing traffic [16]. Therefore, the AID design limits the number of stations that can be associated with an AP to 2007 [16]. Additionally, the Traffic Indication Map (TIM) bitmap enforces the same limit on the number of associated stations as well. The TIM is used for power management mechanisms. It defines the number of buffered frames received from an AP. For these reasons, TGah is extending the range of AID values for 802.11ahs devices from 1-2007 to 0-8191. Also, the IEEE 802.11ah draft standard is increasing the maximal length of the TIM bitmap for 802.11ah devices from 2008 bits to 8192 bits [16]. Therefore, it is quite obvious that ZigBee and IEEE 802.11ah protocols outperform the classic 802.11a/ac protocol when it comes to the network size requirements.

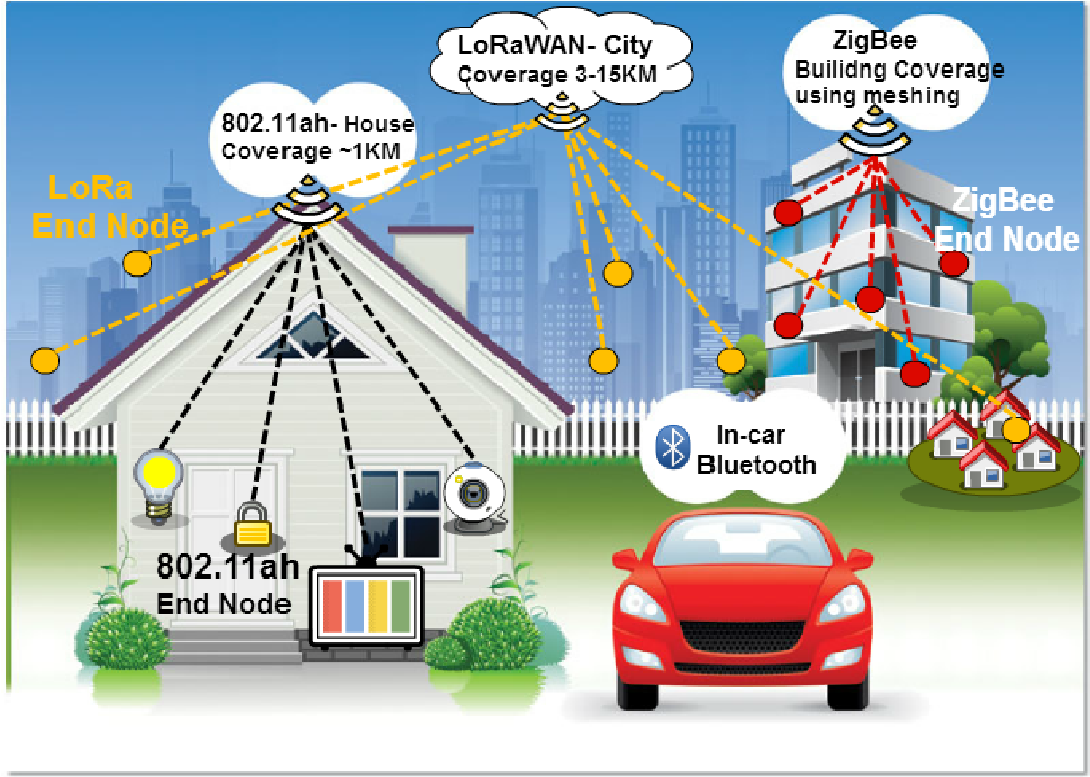

Figure 3- The Influence ofWireless Technologies Ranges on their Applications in the IoT

Table 2- Network Size Comparison

| Technology | Network Size |
|---|---|
| ZigBee | Approximately up to 65,000 nodes |
| Bluetooth | Eight nodes per network/piconet |
| Wi-Fi (802.11a/ac) | 2007 associated with an AP |
| Wi-Fi 802.11ah | Approximately 8000 nodes |
| LoRa | LoRa gateway must have a very high capacity or capability to receive messages from a very high volume of End nodes |

Of course, cellular technologies have an enormous network size. However, cellular connectivity cannot be possible without involving a mobile provider that usually charges a fee per connection. Therefore, while cellular technology can accommodate a larger number of devices, the costs involved are dramatically higher than those associated with other technologies such as ZigBee. LoRa is also expected to support a large volume of devices, however the associated costs, as previously mentioned, are yet to be explored. Table 2 provides a brief comparison between ZigBee, BLE, Wi-Fi, and LoRa with regards to their network sizes.

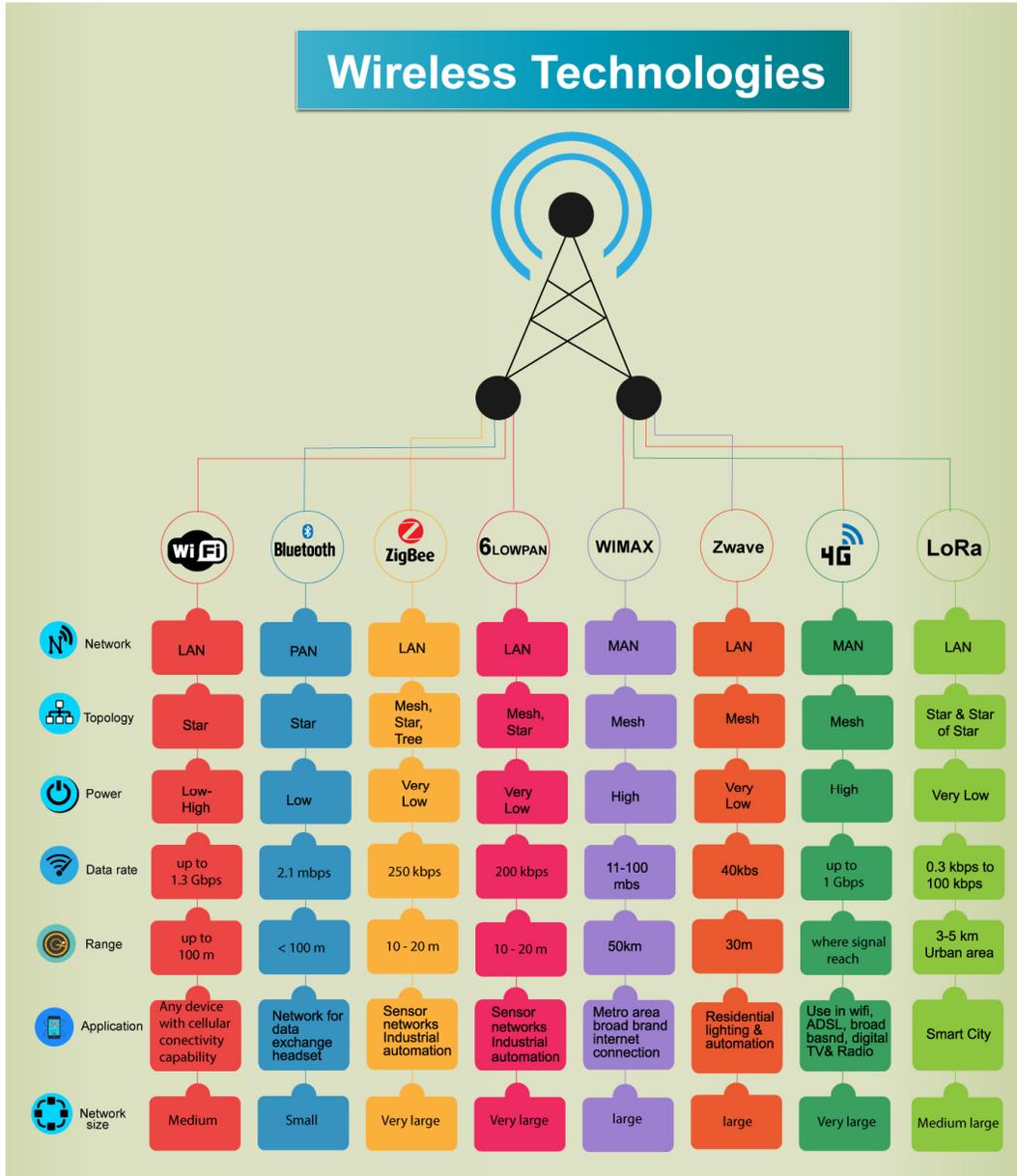

Figure 4- Wireless Technologies Comparison

### 3.3 Transmission Power Evaluation

As shown in Figures 3 and 4, the 802.15.4 based technologies, BLE, LoRa, and 802.11ah all have low power consumption characteristic. The transmission power of BLE ranges from 1 to 10 mW [38]. ZigBee transmission power is very low estimated to be under 1 mW. The Wi-Fi standard has a transmission power of approximately 100 mW. On the other hand, in [39] it was found that with regard to energy consumption, and in the case of a small number of nodes in a low traffic scenario, the IEEE 802.15.4 consumed more average energy for the successful transmission of a packet compared with IEEE 802.11ah. However, in congested networks, the energy consumption of the IEEE 802.11ah was found to be relatively higher than that of IEEE 802.15.4. Therefore, the study concludes that with regard to energy consumption the IEEE 802.15.4 outperformed the IEEE 802.11ah, especially in a dense network with non-saturated traffic characteristics.

Table 3- Low-Power Wireless Technologies Comparison

| | Security | Location Detection | Low Cost | Ease of use | Ecosystem | Range | Remote control | Antenna size | Networking size |
|---|---|---|---|---|---|---|---|---|---|
| 802.15.4 | ✓ | ✓ | ✓ | ✓ | ✓ | ✓(✓) | ✓ | ✓ | ✓ |
| BLE | ✓ | ✗ | ✓ | ✓ | ✓ | ✗ | ✓ | ✓ | ✗ |
| 802.11ah | ✓ | ✓ | ✓ | ✓ | ✓ | ✓ | ✓ | ✓ | ✓(✓) |
| LoRa | ✓ | ✓ | ✓ | ✓ | ✓ | ✓ | ✓ | ✓ | ✓ |

However, when considering the throughput parameter, the IEEE 802.11ah has a better performance when compared to IEEE 802.15.4. Nevertheless, it should be noted that, at the time of writing, the IEEE 802.11ah standard is still under development. Thus, more simulations and experimental studies are required to determine the performance of IEEE 802.11ah effectively. Similarly, the performance and scalability of LoRa over large, dynamic and heterogeneous networks are yet to be explored.

## 4. CONCLUSION

To enable the IoT vision of extending communications to anything and anywhere, the Internet must support connecting things using a variety of wireless and mobile technologies. This paper reviewed some of the enabling wireless technologies in the IoT particularly, ZigBee, 6LoWPAN, BLE, LoRa and Wi-Fi including the low-power IEEE 802.11ah protocol. It examined these technologies and evaluated their capabilities and behaviours with regards to various metrics including the data range and rate, network size, RF channels and bandwidth, power consumption, and the IoT ecosystem. The paper highlighted the unique characteristics of these wireless low-power technologies and the issues about their incorporation in the IoT. It should be noted, however, that the low-power and low-cost characteristics of these technologies and their integration in the IoT demand new management, security, and privacy-preserving methods or approaching the prevailing management and security protection systems differently. There is a need to manage an unprecedented number of things connected to the Internet generating a large amount of traffic across heterogeneous networks, particularly those with low-power capabilities such as those examined in this work. Thus, the challenge remains in supporting secure and interoperable communications between these various technologies creating an ecosystem of coexisted devices rather than isolated islands of networks.

ACKNOWLEDGEMENTS

This research is supported by the International Postgraduate Research Scholarship (IPRS) and the Australian Postgraduate Award (APA).

## AUTHORS

Dr. Mahmoud Elkhodr is with the School of Computing, Engineering and Mathematics at Western Sydney University (Western), Australia. He has been awarded the International Postgraduate Research Scholarship (IPRS) and Australian Postgraduate Award (APA) in 2012-2015. Mahmoud has been awarded the High Achieving Graduate Award in 2011 as well. His research interests include: Internet of Things, e-health, Human Computer-Interactions, Security and Privacy.

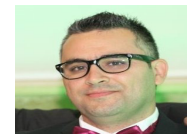

Dr. Seyed Shahrestani completed his PhD degree in Electrical and Information Engineering at the University of Sydney. He joined Western Sydney University (Western) in 1999, where he is currently a Senior Lecturer. He is also the head of the Networking, Security and Cloud Research (NSCR) group at Western. His main teaching and research interests include: computer networking, management and security of networked systems, analysis, control and management of complex systems, artificial intelligence applications, and health ICT. He is also highly active in higher degree research training supervision, with successful results. 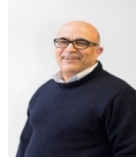

Dr. Hon Cheung graduated from The University of Western Australia in 1984 with First Class Honours in Electrical Engineering. He received his PhD degree from the same university in 1988. He was a lecturer in the Department of Electronic Engineering, Hong Kong Polytechnic from 1988 to 1990. From 1990 to 1999, he was a lecturer in Computer Engineering at Edith Cowan University, Western Australia. He has been a senior lecturer in Computing at Western Sydney University since 2000. Dr Cheung has research experience in a number of areas, including conventional methods in artificial intelligence, fuzzy sets, artificial neural networks, digital signal processing, image processing, network security and forensics, and communications and networking. In the area of teaching, Dr Cheung has experience in development and delivery of a relative large number of subjects in computer science, electrical and electronic engineering, computer engineering and networking. 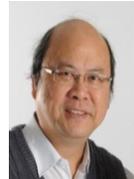